 \documentclass[eps,12pt]{article}
 \usepackage{graphicx}
\begin{document}

\baselineskip=18pt

\begin{center}
{\Large\bf Yang-Lee zeros and the helix-coil transition}
{\Large\bf in a continuum model of polyalanine} 
\vskip 1.1cm
{\bf Nelson A. Alves\footnote{E-mail: alves@quark.ffclrp.usp.br}} 
\vskip 0.1cm
{\it Departamento de F\'{\i}sica e Matem\'atica, FFCLRP 
     Universidade de S\~ao Paulo. \\
      Av. Bandeirantes 3900.  CEP 014040-901 \, 
      Ribeir\~ao Preto, SP, Brazil}
\vskip 0.3cm
{\bf Ulrich H.E. Hansmann \footnote{E-mail: hansmann@mtu.edu}}
\vskip 0.1cm
{\it Department of Physics, Michigan Technological University,
         Houghton, MI 49931-1291, USA}

\vskip 0.3cm
\today
\vskip 0.4cm
\end{center}
\begin{abstract}
We calculate the Yang-Lee zeros for characteristic temperatures of 
the helix-coil transition in a continuum model of polyalanine. 
The distribution of these zeros differs from  predictions of the 
Zimm-Bragg theory and supports recent claims that polyalanine
exhibits a true phase transition. New estimates for critical exponents
are presented and the relation of our results to the Lee-Yang theorem is
discussed.


\end{abstract}
\vskip 0.1cm
{\it PACS-No.: 87.15.Aa, 64.60.Cn, 64.60.Fr }



\section{Introduction} 
\vspace*{-0.5pt}
\indent
  
Recently, there has been an increased interest in the statistical physics of
biological macromolecules. For instance, the conditions under which
\hbox{$\alpha$-helices}, a  common structure in proteins, 
are formed or disolved, have been extensively
studied \cite{Poland}. Traditionary, the characteristics of the 
observed sharp transition between random coil state and (ordered) helix,
have been described in the framework of  Zimm-Bragg-type theories \cite{ZB}.
In these theories, the molecules are  approximated by a one-dimensional
Ising model with the residues as ``spins'' taking values ``helix'' or ``coil'',
and solely local interactions. Hence, thermodynamic phase transitions are
not possible in these theories. 
    However, in previous work 
\cite{HO98c,AH99b,KHC99d} evidence was presented that 
 polyalanine exhibits a phase transition between the ordered 
helical state and the disordered random-coil state when interactions
between all atoms in the molecule are taken into account. 
Investigating the finite size scaling of various
quantities such as specific heat, susceptibility and the Fisher zeros,
we studied the nature of this phase transition and presented estimates for 
the critical exponents $\nu$, $\alpha$ and $\gamma$. 
Here, we  continue our previous investigation by exploring the
helix-coil transition in polyalanine from a different point of view. 
We intend to demonstrate in greater detail
the differences in the critical behaviour of our model and the Zimm-Bragg
theory, and we will present additional  critical exponents.

  Yang and Lee have established long ago \cite{Yang-L} that the statistical
theory of phase transitions can also be described by means
of the distribution of the
zeros of the grand partition function in the complex fugacity  plane.
In the case of the Ising model this description corresponds 
to  study this model with an external magnetic field $H$ 
mathematically extended from real to complex values.
Although complex values for the fugacity are not physical, 
the characteristics of the phase transition  can be extracted from
the distribution of the zeros.
  As the volume of the
given (finite) system increases, the number of complex zeros grows 
and they will close onto the positive real fugacity axis.  
In the thermodynamic limit, those zeros 
circumscribe closed regions on this plane, thus defining thermodynamic 
phases which are themselves free of zeros. 
In Ref.~\cite{Lee-Y} Lee and Yang  proved 
the famous circle theorem, namely that in the Ising model with 
ferromagnetic couplings the complex zeros lie on the unit circle
in the complex $y-$plane,  $y = e^{-2 \beta H}$.

The goal of our paper is to investigate whether the helix-coil transition
in biological macromoelecules can also be described in the frame work
of the Yang-Lee zeros.  For polyalanine, the analogous of the magnetization
in the Ising model is  the  number of helical residues $M$. 
Formally we can  introduce a (non physical) 
external field $H$ as the conjugate variable to this
order parameter $M$ (which plays the role of a magnetization), and study the
Yang-Lee zeros. Such an investigation allows us not only to check our 
previous results with an independent method, it opens also the
possibility to calculate new critical exponents and to point out 
much clearer the differences between our all-atom model of polyalanine
and the Zimm-Bragg theory. If the helix-coil transition can indeed be described
by the Zimm-Bragg model, then the distribution of the Yang-Lee zeros 
in function of the temperature
should
resemble that of the one dimensional Ising model. On the other hand, a 
substantially different distribution would demonstrate that the helix-coil
transition in polyalanine is not accurately 
described by the Zimm-Bragg theory.

Before proceeding, we give the outline of the paper. In the next
section we describe the numerical evaluation of the partition function.
Our approach and numerical results are presented in the third Section, 
which is followed by our conclusions.

\section{Methods}
\noindent
Our investigation of the helix-coil transition for polyalanine is
based on a detailed, all-atom representation of that homopolymer. 
Since one can avoid the complications of electrostatic and 
hydrogen-bond interactions of side chains with the solvent for alanine 
(a nonpolar amino acid), explicit solvent molecules were neglected. 
The interaction between the atoms  was
described by a standard force field, ECEPP/2,\cite{EC}  (as implemented 
in the KONF90 program \cite{Konf}) and is given by:
\begin{eqnarray}
E_{tot} & = & E_{C} + E_{LJ} + E_{HB} + E_{tor},\\
E_{C}  & = & \sum_{(i,j)} \frac{332q_i q_j}{\epsilon r_{ij}},\\
E_{LJ} & = & \sum_{(i,j)} \left( \frac{A_{ij}}{r^{12}_{ij}}
                                - \frac{B_{ij}}{r^6_{ij}} \right),\\
E_{HB}  & = & \sum_{(i,j)} \left( \frac{C_{ij}}{r^{12}_{ij}}
                                - \frac{D_{ij}}{r^{10}_{ij}} \right),\\
E_{tor}& = & \sum_l U_l \left( 1 \pm \cos (n_l \chi_l ) \right).
\end{eqnarray}
Here, $r_{ij}$ (in \AA) is the distance between the atoms $i$ and $j$, and
$\chi_l$ is the $l$-th torsion angle. Note that with the electrostatic
energy term $E_{C}$ our model contains a long range interaction neglected
in the Zimm-Bragg theory \cite{ZB}.  It was  conjectured  that it is 
this long range interaction in our model and  the fact that it is not  
one-dimensional which allows the existence of the observed phase
transition in Refs.~\cite{HO98c,AH99b,KHC99d}. 
We remark that  the 1D Ising model with long-range interactions
also exhibits  a phase transition at finite $T$ if the interactions
decay like $1/r^{\sigma}$ with $1\le \sigma < 2$ \cite{1dlr}.

Simulations of detailed models of biological macromolecules are 
notoriously difficult. This is because the various competing interactions 
within the polymer lead to an energy landscape characterized by a 
multitude of local minima.  Hence, in the low-temperature region, 
canonical Monte Carlo or molecular dynamics 
simulations will tend to get trapped in one of these
minima and the simulation will not thermalize within the available
CPU time. Only recently, with the introduction of new and sophisticated 
algorithms such as  {\it multicanonical sampling} \cite{MU} and other 
{\it generalized-ensemble} techniques \cite{Review} was it
possible to alleviate this problem in  protein simulations \cite{HO}. 
For polyalanine,  both the failure of standard Monte Carlo techniques 
and the superior performance of the multicanonical algorithm are 
extensively documented  in earlier work \cite{OH95b}. For this reason,
we use again this simulation technique for our project  where we
considered polyalanine chains of up to $N=30$ monomers. 
 
In the multicanonical algorithm \cite{MU}
conformations with energy $E$ are assigned a weight
$  w_{mu} (E)\propto 1/n(E)$. Here, $n(E)$ is the density of states.
A  simulation with this weight
will  lead to a uniform distribution of energy:
\begin{equation}
  P_{mu}(E) \,  \propto \,  n(E)~w_{mu}(E) = {\rm const}~.
\label{eqmu}
\end{equation}
This is because the simulation generates a 1D random walk in the 
energy space,
allowing itself to escape from any  local minimum.
Since a large range of energies are sampled, one can
use the reweighting techniques \cite{FS} to  calculate thermodynamic
quantities over a wide range of temperatures $T$ by
\begin{equation}
<{\cal{A}}>_T ~=~ \frac{\displaystyle{\int dx~{\cal{A}}(x)~w^{-1}(E(x))~
                 e^{-\beta E(x)}}}
              {\displaystyle{\int dx~w^{-1}(E(x))~e^{-\beta E(x)}}}~,
\label{eqrw}
\end{equation}
where $x$ stands for configurations.

Note that unlike in the case of canonical simulations the weights are
not a priori known. 
We needed between 40,000 sweeps ($N=10$)  and 500,000 sweeps ($N=30$) for
the weight factor calculations by the iterative  procedure described in 
Refs.~\cite{MU,HO94c}. 
All thermodynamic quantities were  estimated from one
production run of $N_{sw}$ Monte Carlo sweeps starting from a random
initial conformation, i.e. without introducing any bias.
We chose $N_{sw}$=2\,000\,000, 2\,000\,000, 4\,000\,000, and 3\,000\,000 
sweeps for $N=10$, 15, 20, and 30, respectively.

It follows from Eq.~(\ref{eqmu}) that the multicanonical algorithm allows us
to calculate estimates for the spectral density:
\begin{equation}
  n(E) = P_{mu} (E) w^{-1}_{mu} (E)~.
\end{equation}
We can therefore construct
the corresponding partition function for our all-atom model of polyalanine
keeping track of the corresponding number of helical residues $M$ at a 
given energy,  
\begin{equation}
Z(u,y)\,=\, \sum_{M=0}^{N-2} \sum_E n(E,M) u^E y^M ,     \label{eq:z1}
\end{equation}
where 
$u = e^{-\beta}$ with $\beta$ the inverse temperature,
$\beta=1/k_BT$ and $y=e^{-H}$. 
We have absorbed the factor $\beta$ in the 
definition of the  field $H$ which is the conjugate variable to our
order parameter $M$. The later
quantity is defined  by the condition that for a helical residue 
the dihedral angles ($\phi,\psi$) fall in the
range ($-70 \pm 20^{\circ},-37 \pm 20^{\circ}$).
Note, that for convenience we have discretized the energy in Eq.~(\ref{eq:z1}).
In previous work 
we found that the partition function zeros depend only
weakly on the energy bin size, and for the present study we choose 
energy bins of length $0.5$ kcal/mol.

\section{Yang-Lee Zeros of Polyalanine}
\noindent
In order to analyse the helix-coil transition in our all-atom model
of polyalanine we  first determine the (pseudo)critical temperatures.
This can be done in various ways. For instance, in Ref.~\cite{AH99b}
these temperatures where calculated from the distribution of the
Fisher zeros \cite{Itzykson}. Here, we  check our previous results
by calculating estimates for the pseudo-critical temperatures in a 
different way.
  
 Our temperature-driven transition model can be analysed by considering
the coexistence of two phases: a disordered (coil) phase  and
an ordered (helix) phase at the critical temperature $T_c$.
This behaviour can be described by the existence of an equilibrium
probability distribution $P_N(M)$ for the corresponding order 
parameter $M$ \cite{BinderB30,ChallaB34}. 
  This approach leads to a definition of a finite size critical temperature
by approximating the equilibrium distribution by a sum of two
distributions, each one characterizing the corresponding phase.
  This ansatz gives origin  to a two-peak like distribution separated by a
minimum and has led to the finite size scaling study
of the coexistence of bulk phases \cite{Lee-K90}.
  Following these works, we consider for our  purpose
the histograms of the helicity for all chain
sizes $N$, which is our order parameter (and 
the equivalent to the magnetization in the Ising model), 
at different temperatures $\beta$. 
   It is straightforward to calculate the histograms of 
helicity distribution at these temperatures
from our multicanonical estimates of the density of states,
\begin{equation}
                     w(M)\,=\, \sum_E n(E,M) u^E .     \label{eq:hist}
\end{equation}
   As expected,
these histograms exhibit a clear double peak structure at 
what we call critical temperature, while at high (low) temperatures one 
observes solely a predominant single peak at low (high) values of the 
helicity. 
  Hence, we define our critical temperatures by the condition that
the reweighted histogram of the magnetization have two peaks of equal heights.
  As an example we show in Fig.~1 the histogram normalized to area one,
for $N=30$ at $T_c=518$ K.

  Table 1 summarizes the so obtained set of critical temperatures by this
method. The quoted values and errors were obtained by reweighting 
independent binned data to temperatures $T_c^i$, $i=1,2, ..., nbins$, 
where one obtains equal heights in the histograms for our order 
parameter $M$.  Our statistics were based on $nbins = 19, 20, 
16 \,{\rm and}\, 3$ for  $N=10,15, 20\, {\rm and}\, 30$, respectively. 
for comparison we also list our previous estimates \cite{AH99b}
which were derived from an analysis of the Fisher zeros and the position
of the maximum of specific heat $C_v$.  We note, that our new results 
agree within the errors with our previous estimates for the 
pseudo-critical temperatures which demonstrates the validity
of our approach.

After having demonstrated the reliability of our data we calculate now 
at the so obtained critical temperatures the distribution of the Yang-Lee 
zeros by means of Eq.~(\ref{eq:z1}).  Our results are displayed in 
Fig.~2 for the case of chain lengths $N=30$. In that figure
 we draw in addition a  
unit circle to point out the relative position of our zeros around it. 
  Similar distributions are obtained for smaller systems.
  The presented error bars for the zeros in this figure, 
were estimated by considering the new partition functions  at 
the limiting temperatures  $T_c(N) \pm \Delta T_c(N)$ where $\Delta T_c(N)$
is the error in our pseudo-critical temperatures $T_c(N)$. 
  Our error bars  clearly 
exclude the possibility that the complex zeros are distributed
on the unit circle. 
  On the other hand, the Yang-Lee zeros for the 
1D Ising model (which can be calculated exactly) fall on the unit circle. 
This is shown in Fig.~3 for chain lengths 
$N=5, 15$ and $30$.\footnote{As a further illustration for the  behaviour 
      of the continuous loci of zeros for the one dimensional 
      Ising partition function, we refers to Ref.~\cite{Katsura},
      and Ref.~\cite{Glumac94} for $q=2$ states.} 

After having demonstrated the difference in the two models
at the critical point, we extend our study to the behaviour 
of our polyalanine
zeros as function of the temperature.
  In Fig. 4 we show for our smallest polyalanine chain how the zeros 
change with temperature $u(\beta)$, measured in  multiple 
of the critical value $u_c$. Again, we draw for illustration a unit circle
to call attention to the relative position of our zeros. 
  We exhibit the smallest chain since the density
of zeros increases with growing chain length making it difficult to
the eyes to
trace the movement of the zeros in the complex $y$-plane. 
 As the temperature increases,  $u(\beta) > u_c$, 
we observe that the zeros move away from the unit circle and
consequently the edge zero, the zero closest to the positive real $y-$axis,
moves away from the real axis.
     A similar behaviour is known for the $q$-state 
     Potts model with $q>2$ on finite lattices for temperatures 
     larger than the critical one \cite{Kim-Cres1}.
 For $u(\beta) < u_c$, we find again that
the zeros do not fall all on the unit circle. 
Hence, the Lee-Yang theorem does not apply for our model demonstrating again 
that polyalanine should not be described by  Ising-type models. 

 Figure 4 indicates that the angle $\theta_0$, the angle at
which the  edge zeros deviate from the real $y$-axis, grows as 
the temperature is increased  above $T_c$, i.e. $u(\beta) >  u_c$. 
 For larger lengths $N$, the number of zeros increases and the
edge zeros move towards the real axis. 
Note  that the movement for the edge zero seems to follow an
unit arc, leading to $Re(y)=1$ and  the real value $H=0$ in the 
thermodynamic limit.  The movement of the edge zeros towards the 
real axis itself is a signal for the 
possibility of a true phase transition in the thermodynamic limit
if the locus of zeros cut the positive real axis at finite temperature.
In that case one would expect the following scaling relation for the 
edge zeros \cite{Kim-Cres2}:
\begin{equation}
\theta_0(t, N) \,=\, N^{-y_h/d} \theta_0(t N^{y_t/d})\, ,   \label{eq:t0}
\end{equation}
where $t= (T-T_c)/T_c$, $y_h$ is the magnetic scaling exponent
and $y_t = 1/\nu$ is the thermal scaling exponent.
Note, that we have introduced the scaling variable $N^{1/d}$ 
in place of the usual linear length $L$. This is because the number
of monomers $N$ is the natural quantity to describe polyalanine chains.
  Since we have no theoretical indication to assume  a particular
integer geometrical dimension $d$ for our polypeptide, we present our
results as a function of this parameter $d$.

Now, assuming that the above finite size scaling relation holds for polyalanine
chains we can calculate the ``magnetic'' scaling exponent \cite{Itzykson}
\begin{equation}
                  y_h/d = (\beta + \gamma)/d\nu
\label{eq:yhb}
\end{equation}
 if we take $t=0$, i.e.  
$u(\beta) = u_c$. For this reason
we display in Fig. 5 the linear regression for ${\rm ln}\,(\theta_0(N))$. 
 The least-square fit gives us
\begin{equation}
                    y_h/d = 0.97(4) \, ,             \label{eq:hy}
\end{equation}
with the goodness of fit $Q=0.67$. This value for $Q$ means
the fit is acceptable
for the present precision data and corroborates our expectation.


If we take our estimate in Eq.~(\ref{eq:hy}) and the reported values
for $\gamma$ and $d\nu$ from Ref.~\cite{AH99b}, $\gamma= 1.06(14)$,
$d\nu = 0.93(5)$, we can evaluate the exponent $\beta$ either 
by the defining equation Eq.~(\ref{eq:yhb}) or from the relation
$ 2 \beta \,+\, \gamma = d \nu$. The first option is less precise since
it envolves three parameters 
($\beta=-0.16(15)$),  while the second one, relying on two previous estimates,
 gives $\beta=-0.065(74)$. As one should expect, both estimates agree
with each other within the error bars demonstrating the consistency of
this study with earlier investigations \cite{HO98c,AH99b,KHC99d}. 
Note also that our estimates for the critical exponent $\beta$ do not
help us to answer a question we had  to leave unanswered in
the earlier work, namely, whether the helix-coil transition is a
first order phase transition ($\beta = 0$) or a (strong) second order
phase transition ($\beta \ne 0$). This is not surprising since clarifying 
the  nature of a phase transition can be a challenging issue 
even for
simpler systems (such as Potts models \cite{LandauB39}, 
to name only one example)
and requires dedicated large-scale simulations.
The important point here is  rather that our model of polyalanine, 
despite its complexity, can be described in terms of a critical theory.

\section{Conclusion}
\indent
We have extended the Yang-Lee zeros analysis to an all-atom model 
of polyalanine polymers of various chain lengths. 
 Our results were compared with 
the Zimm-Bragg model which is often used to describe helix-coil
transitions in Biopolymers.
  Our analyses of the Yang-Lee zeros for both models 
  give clear evidence for our claim that the helix-coil 
  transition in an all-atom model of polyalanine is not 
  adequately described by the Zimm-Bragg theory.
 Our results confirm
earlier work where  we pointed out the possibility of a thermodynamic
phase transition in polyalanine which is forbidden in the Zimm-Bragg model.

\vspace{0.4cm}
{\bf Acknowledgements}
\vspace{0.4cm}

 U. Hansmann gratefully acknowledges support by  a research grant 
of the National Science Foundation (CHE-9981874). This article was
finished while U.H. was visitor at the Department of Physics at
the Bielefeld University. He thanks F.~Karsch for his hospitality
during his stay in Bielefeld.



\newpage
{\huge Table:}\\

\noindent
\begin{table}[ht]
\renewcommand{\tablename}{Table}
\caption{\baselineskip=0.8cm  Critical temperatures $T^{hist}_c$ as
obtained by our analysis for polyalanine chain lengths $N$. For comparison,
we show also previous estimates obtained from both an analysis of
Fisher zeros ($T_c^{\rm Fisher}$) and the location of the peak in
the specific heat ($T_c^{C_v}$).}
\begin{center}
\begin{tabular}{l c c c c }\\
\\[-0.3cm]
~~$N$  &~~$T_c^{\rm \,hist}$  
      &~~$T_c^{\rm \,Fisher}$   &~~$T_c^{\, C_v}$ \\
\\[-0.35cm]
\hline
\\[-0.3cm]
~~10 &  ~$433(1)$  & ~442(8)  & ~427(7) \\
~~15 &  ~$496(3)$  & ~498(4)  & ~492(5) \\
~~20 &  ~$518(4)$  & ~511(5)  & ~508(5) \\
~~30 &  ~$518(3)$  & ~520(4)  & ~518(7) \\
\end{tabular}
\end{center}
\end{table}
\vfill

\newpage

{\huge Figure Captions:} \\

{\bf Figure 1.} Histogram for the number of helical residues in conformations
for chain length $N=30$. \\

{\bf Figure 2.} Distribution of complex partition function zeros in the
$y-$plane with error bars, at $T_c(N=30)$. 
 We draw an inexistent unit circle to show
how the zeros are distributed around it.
 The zeros do not circumscribe an unit circle as expected
for the Lee-Yang theorem in Ising like systems with positive interactions. \\

{\bf Figure 3.} The Yang-Lee zeros for the Ising chain with lengths 
$N=5, 15$ and 30. \\

{\bf Figure 4.} The behaviour of the partition function zeros for polyalanine
with chain length $N=10$ in function of the temperature $u(\beta)$. \\

{\bf Figure 5.} Linear regression for ${\rm ln}\,(\theta_0)$,
       Eq.~(\ref{eq:t0}), at the critical temperature. 
\vfill


\newpage

\begin{figure}[t]
\begin{center}
\begin{minipage}[t]{0.95\textwidth}
\centering
\includegraphics[angle=-90,width=0.72\textwidth]{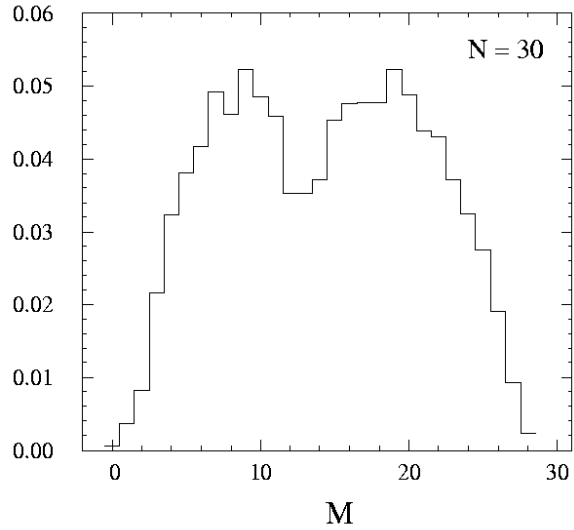}
\renewcommand{\figurename}{Fig.}
\caption{Histogram for the number of helical residues in conformations
for chain length $N=30$.}
\label{fig1}
\end{minipage}
\end{center}
\end{figure}
\vfill
\newpage

\begin{figure}[!ht]
\begin{center}
\begin{minipage}[t]{0.95\textwidth}
\centering
\includegraphics[angle=-90,width=0.72\textwidth]{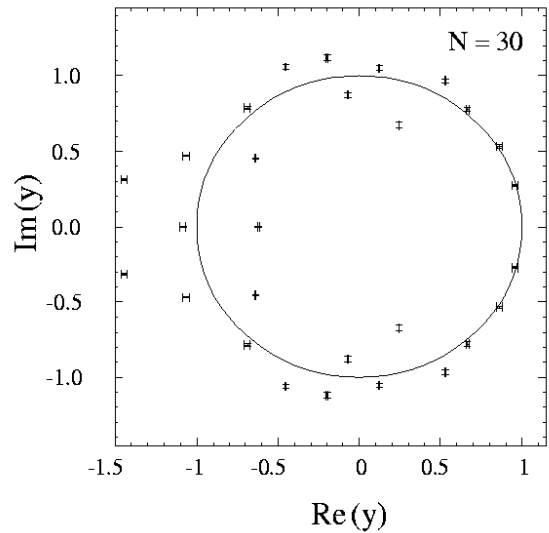}
\renewcommand{\figurename}{Fig.}
\caption{Distribution of complex partition function zeros in the
$y-$plane with error bars, at $T_c(N=30)$. 
 We draw an inexistent unit circle to show
how the zeros are distributed around it.
 The zeros do not circumscribe an unit circle as expected
for the Lee-Yang theorem in Ising like systems with positive interactions.}
\label{fig2}
\end{minipage}
\end{center}
\end{figure}
\vfill

\newpage
\cleardoublepage

\begin{figure}[t]
\begin{center}
\begin{minipage}[t]{0.95\textwidth}
\centering
\includegraphics[angle=-90,width=0.72\textwidth]{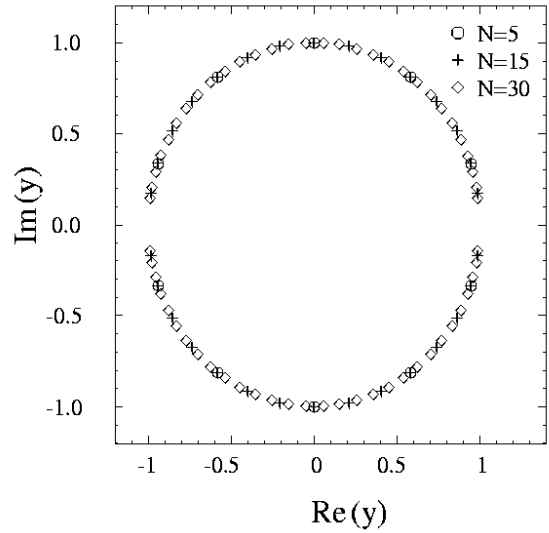}
\renewcommand{\figurename}{Fig.}
\caption{The Yang-Lee zeros for the Ising chain with lengths 
$N=5, 15$ and 30.}
\label{fig3}
\end{minipage}
\end{center}
\end{figure}
\vfill

\newpage

\begin{figure}[!ht]
\begin{center}
\begin{minipage}[t]{0.95\textwidth}
\centering
\includegraphics[angle=-90,width=0.72\textwidth]{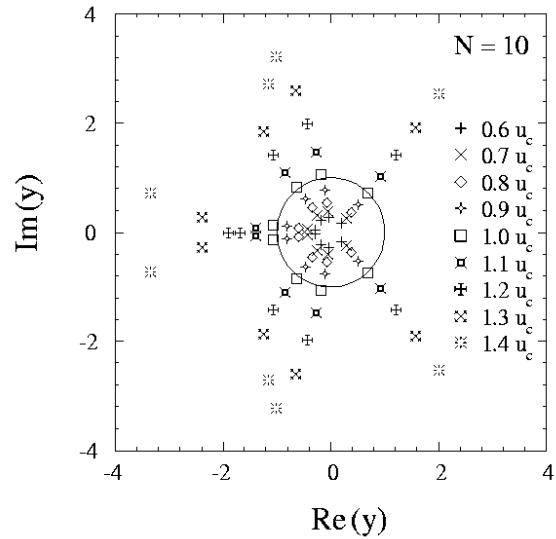}
\renewcommand{\figurename}{Fig.}
\caption{The behaviour of the partition function zeros for polyalanine
with chain length $N=10$ in function of the temperature $u(\beta)$.}
\label{fig4}
\end{minipage}
\end{center}
\end{figure}
\vfill

\newpage
\cleardoublepage

\begin{figure}[!ht]
\begin{center}
\begin{minipage}[t]{0.95\textwidth}
\centering
\includegraphics[angle=-90,width=0.72\textwidth]{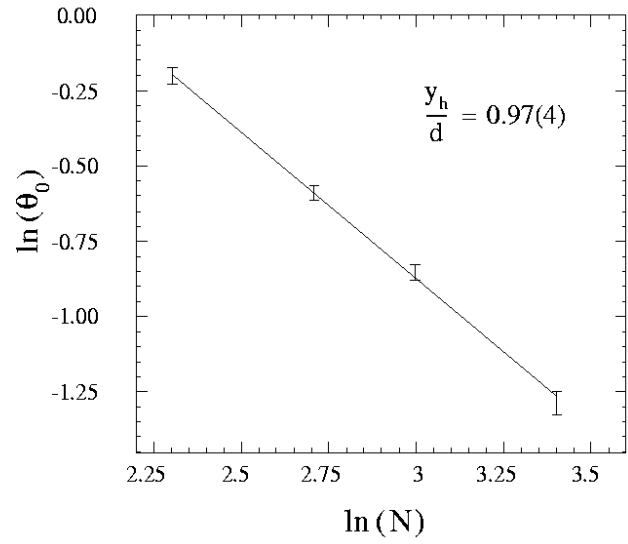}
\renewcommand{\figurename}{Fig.}
\caption{Linear regression for ${\rm ln}\,(\theta_0)$,
       Eq.~(\ref{eq:t0}), at the critical temperature.}
\label{fig5}
\end{minipage}
\end{center}
\end{figure}
\vfill

\end{document}